# Toward catalyst-mediated growth of diamond and cubic boron nitride wires

Perspective

Rodney S. Ruoff[1,2,3,4,*]

[1] Center for Multidimensional Carbon Materials (CMCM), Institute for Basic Science (IBS), Ulsan 44919, Republic of Korea
[2] Department of Chemistry, UNIST, Ulsan 44919, Republic of Korea
[3] Department of Materials Science and Engineering, UNIST, Ulsan 44919, Republic of Korea
[4] School of Energy and Chemical Engineering, UNIST, Ulsan 44919, Republic of Korea
[*] Corresponding author: ruofflab@gmail.com

Running title: Diamond and cBN wire growth

*Acknowledgments and funding*

This work was supported by the Institute for Basic Science (IBS-R019-D1). Rodney S. Ruoff conceived the Perspective, assessed the literature, and wrote and revised the manuscript, with the assistance disclosed below.

*Data availability statement*

No new experiments or atomistic simulations were performed. Numerical estimates, input data and assumptions are provided in the Supporting Information; a reproducible calculation script accompanies the revision. All other data discussed are from the cited literature.

*Conflict of interest*

The author declares no conflict of interest.

Abstract

Sustained catalyst-mediated growth of diamond and cubic boron nitride (cBN) wires has yet to be established reproducibly, although related methods produce carbon and BN nanotubes and many elemental and compound nanowires. I hypothesize that a retained contact between a catalyst and an existing tetrahedral crystal can sustain wire extension under conditions that do not nucleate that crystal on an otherwise comparable unseeded catalyst. The prediction is repeated, vapor-derived target-phase extension during an interval in which matched unseeded contacts do not nucleate that phase. I develop tests using diamond seeds, SiC starter wires and cBN seeds, separating the conditions used to establish the contact from those used for continuation. Worked estimates distinguish bulk precipitation driving forces from sidewall costs and show why increased carbon activity alone cannot select diamond over graphite. Comparisons with established synthesis identify the roles of pressure, hydrogen and activated nitrogen that must be retained or replaced. Isotope tracing, local structural measurements and time-dependent growth distinguish continued incorporation from seed consumption, conversion and etching. Nanometer, micrometer and larger wires could provide electrically insulating thermal connections, mechanical probes and reinforcing filaments.

## 1 The synthesis problem

Catalyst-mediated growth has produced Si and Ge wires, compound wires such as SiC, GaAs and GaN, and alloys such as SiGe and InGaAs. [1, 2, 3] Carbon and boron nitride are readily obtained as nanotubes with three-coordinate bonding. [4, 5] Reproducible sustained growth of diamond or cubic boron nitride (cBN) wires at exposed catalyst contacts remains unresolved. Early diamond-whisker reports by Derjaguin and colleagues provide an experimental starting point, but their gas composition, catalyst state and contribution from the diamond substrate were not fully resolved. [6, 7]

In vapor–liquid–solid (VLS) growth, vapor supplies material to a liquid particle at the growth contact; in vapor–solid–solid (VSS) growth, the particle is solid. [8, 9] Our collaboration with the Buhro group produced crystalline BN nanotubes using nickel-boride particles and proposed incorporation at the root, without establishing whether those particles were liquid during growth. [10] Santra and colleagues reported BN nanotube growth at liquid boron surfaces without added transition metals. [11]

Our Cu-assisted hydrogenated carbon nanofibers were amorphous despite their substantial $sp^3$ bonding. [12] Reported diamond nanowires grown inside nanotubes used a confining structure; the experiments proposed here seek continued growth at an exposed catalyst contact. [13, 14]

I propose separating the preparation of a suitable catalyst–crystal contact from its subsequent operation. An existing diamond or cBN seed supplies the target lattice. *A* starter wire of another material supplies a small, retained contact at which the target must first nucleate. These are different tests: seeded continuation bypasses heteronucleation, whereas a starter switch tests whether the new phase can establish itself before the original contact dissolves or changes shape.

The central hypothesis is that an established tetrahedral crystal–catalyst contact has a continuation window in which repeated target-phase growth occurs although an otherwise comparable unseeded catalyst does not nucleate that phase. I would compare seeded and unseeded contacts at matched catalyst composition, footprint and vapor exposure, and then switch established contacts into the same continuation conditions. The predicted result is repeated vapor-derived extension during a defined observation interval in which matched unseeded contacts do not form the target phase. Comparable target formation in both groups would not support the proposed advantage under those conditions. Loss of the target structure after a finite initial deposit would fail the continuation test.

### 1.1 Properties and applications

Diamond and cBN wires could serve as small mechanical beams, wear-resistant probes or reinforcing filaments. Diamond cleaves preferentially on {111}, so orientation and surface flaws will matter to tensile failure. Large reversible bending strains have nevertheless been reported for fabricated diamond nanoneedles. [15, 16] Bu and colleagues reported orientation-dependent deformation of cBN pillars under compression. [17] The pillar compression results do not establish tensile strength or crack resistance in long wires.

High thermal conductivity combined with electrical insulation could allow a wire to connect a hot component to a heat sink without an electrical short. Anthony and colleagues reported about 2200 W $m^{-1}$ $K^{-1}$ for natural-abundance diamond and 3300 ± 200 W $m^{-1}$ $K^{-1}$ for approximately 99.9% $^{12}C$ diamond at 25 °C. [18] Chen and colleagues reported room-temperature conductivity above 1600 W $m^{-1}$ $K^{-1}$ in isotope-enriched cBN crystals. [19] Wire dimensions, surfaces, grain boundaries and end contacts can reduce heat transfer relative to these crystal benchmarks. [20, 21] Color centers in diamond nanostructures and room-temperature optically detected magnetic resonance from defect ensembles in cBN also motivate scanning probes for local optical or magnetic measurements. [22, 23]

Both single-crystal and polycrystalline wires merit investigation, including micrometer and larger diameters. Here wire means an elongated solid and does not require a circular cross-section. Wurtzite BN and hexagonal diamond are additional tetrahedral targets with different stacking sequences. [24, 25] Layered hexagonal BN is denoted hBN; it is distinct from wurtzite BN.

## 2 What established synthesis provides

*A* proposed contact must provide a route for precursor activation, target-phase incorporation and removal or suppression of competing products. Established diamond and cBN methods accomplish these tasks in different ways (Table 1). High pressure can favor a dense phase; a solvent transports material and changes its activity; reactive hydrogen or energetic species can alter surface termination and competing deposition. *A* seed supplies the first target lattice; new layers must still nucleate or advance from existing steps.

Table 1. Established synthesis and the functions relevant to proposed wire growth.

| Established route | Representative conditions and function | Consequence for wire proposals |
|---|---|---|
| Diamond HPHT | Fe-, Ni- or Co-containing solvents at several GPa in the diamond-stable region transport carbon. Pressure changes the bulk phase balance. [26] | *A* low-pressure Ni contact does not inherit the HPHT phase stability. |
| Diamond CVD | Hot-filament example: filament ≈2000 °C, substrate ≈900 °C, $CH_4$:$H_2$ = 1:100 and ≈4000 Pa. Activated hydrogen participates in surface chemistry. [27] | *A* plasma is not required to generate atomic H. *A* proposed metal contact must still supply activation and control competing carbon. |
| Diamond in liquid metal | Ga/Ni/Fe/Si = 77.75/11/11/0.25 at.%; local growth region ≈1025 °C; $CH_4$/$H_2$ = 5/100 sccm; 1 atm. [26] | Diamond formed without an external plasma; $H_2$ was present. Thickness did not continue increasing over longer runs. |
| cBN HPHT solution growth | $Li_3BN_2$ solvent; 5.5–6 GPa, 1700–1950 °C and 15–30 *h*; source-to-seed transport. [28] | *A* seeded-growth control, outside vapor-fed VLS/VSS. Seed presence does not eliminate dissolution. |
| cBN ion-assisted deposition | On (001) diamond at 900 °C, with 280 eV Ar/N assisting ions. [29] | *A* demonstrated cubic film route. *A* metal-covered seed without bombardment remains to be tested. |

Our group reported diamond formation in Ga–Ni–Fe–Si liquid exposed to methane and hydrogen at 1 atm, with the growth region near 1025 °C. [26] Carbon entered from the gas, but film thickness stopped increasing during longer growth runs. This experiment provides one reference condition for testing a confined metal–diamond contact. It establishes neither continuous wire growth nor a preferred temperature for other catalyst compositions or contacts.

I do not assume that atomic hydrogen, plasma or ions are unnecessary. *A* thermally operated catalyst may activate molecular hydrogen, and a remote nitrogen plasma could provide reactive nitrogen without intentionally bombarding the buried growth interface. Comparisons should determine whether the desired crystal grows because the contact favors incorporation, because activated species suppress a competing phase, or because both are required. The relevant comparison for cBN includes HPHT seeded growth as well as ion-assisted film deposition.

## 3 Thermodynamic considerations

### 3.1 Bulk driving forces and the catalyst

For a locally equilibrated catalyst, the chemical potential of a component is the partial derivative of total Gibbs energy with respect to its amount at fixed temperature, pressure and amounts of the other components. Per mole, it is the partial molar Gibbs energy; here it is expressed per atom. For a pure homogeneous bulk elemental phase, $\mu$ equals Gibbs energy per atom. For stoichiometric bulk BN in equilibrium with B and N reservoirs, $\mu_B + \mu_N$ equals $g_{BN}$ per formula unit. Surface and interface contributions to a finite particle must be included separately when evaluating the derivative. The bulk precipitation driving forces used here are

$$\begin{aligned} \Delta\mu_{D} &= \mu_{C}^{cat} - g_{D} \\ \Delta\mu_{SiC} &= \mu_{Si}^{cat} + \mu_{C}^{cat} - g_{SiC} \\ \Delta\mu_{cBN} &= \mu_{B}^{cat} + \mu_{N}^{cat} - g_{cBN} \end{aligned} \tag{1}$$

The superscript cat denotes the catalyst. Elemental chemical potentials $\mu$ are per atom; $g_D$ is per carbon atom and $g_{SiC}$ and $g_{cBN}$ are per formula unit. Positive $\Delta\mu$ favors precipitation before finite-size, interface and strain contributions are added. Catalyst composition enters through $\mu_i^{cat} = \mu_i^{ref} + k_B T \ln a_i$, where $a_i$ is the activity relative to the stated reference. *A* measured concentration does not determine that activity without a thermodynamic description of the alloy.

$$\begin{aligned} \Delta\mu_{graphite} &= \mu_C^{cat} - g_{graphite} \\ \Delta\mu_{hBN} &= \mu_B^{cat} + \mu_N^{cat} - g_{hBN} \end{aligned}$$

For graphite and hBN, Eq. 1 is extended by replacing the target-solid free energy with that of the competing phase, as written above. At a common carbon reservoir, $\Delta\mu_D - \Delta\mu_{graphite} = g_{graphite} - g_D$.

The corresponding BN difference is $g_{\mathrm{hBN}} - g_{\mathrm{cBN}}$. Increasing the reservoir chemical potential increases the bulk driving force for both phases of the same composition; it does not reverse their bulk ordering. Published BN thermodynamic assessments disagree on that ordering in parts of the temperature–pressure range, so a numerical comparison should retain the alternative free-energy descriptions. [30, 31]

Catalysis can change precursor activation and attachment barriers without changing the bulk free-energy difference between fixed pure phases. *A* finite catalyst also creates interfaces, changes solute activities and can impose strain. Those contributions can change the free energy of the entire junction and the barriers to nucleating or extending either structure. Eichhammer and colleagues combined alloy thermodynamics, segregation and interface terms for finite catalyst–wire systems. [32] Applying that treatment to diamond or cBN requires the relevant competing carbide, boride, nitride and three-coordinate phases.

### 3.2 Diameter and sidewall formation

Extension creates new sidewall area. For a wire of fixed cross-sectional area *A*, facet perimeter lengths $\ell_j$, sidewall free energies $\gamma_j$ and volume $\Omega$ per incorporated atom or formula unit, the added sidewall cost is

$$\Delta\mu_{\mathrm{side}} = \frac{\Omega}{A}\sum_j \gamma_j \ell_j = \frac{2\gamma_{\mathrm{side}}\Omega}{r} \qquad (2)$$

The second equality applies to a circular wire of radius $r$ with isotropic $\gamma_{\mathrm{side}}$ and fixed surface chemistry. *A* large seed may carry a small catalyst deposit, so seed size, catalyst size and wire diameter are independent experimental variables. The sidewall term concerns extension; catalyst curvature, segregation, changes in contact area and a new nucleus require additional terms.

Using representative lattice parameters gives $\Omega = 5.67 \times 10^{-30}$ m$^3$ per C atom and $1.18 \times 10^{-29}$ m$^3$ per BN pair. [33, 19] At a diameter of 10 nm, $\gamma_{\mathrm{side}} = 1$ J m$^{-2}$ gives sidewall costs of 0.014 eV per C atom and 0.030 eV per BN pair; at 1 μm they are 100 times smaller. Taking $\gamma_{\mathrm{side}} = 0.1$–$5$ J m$^{-2}$ as a sensitivity range gives 0.0014–0.071 eV per C atom and 0.0030–0.15 eV per BN pair at 10 nm. These are selected surface-energy values, not measurements or predictions for the proposed metal contacts.

For a 10 nm diamond wire, the sidewall terms are about 1.4 and 6.8 kJ mol$^{-1}$ at $\gamma_{side}$ = 1 and 5 J m$^{-2}$, respectively. The selected diamond–graphite bulk difference at 1200 K is 6.8 kJ mol$^{-1}$. [34] They cannot establish a universal diameter at which diamond or cBN becomes preferred, because the competing structure also creates surfaces and interfaces. Barnard and Snook's calculation of hydrogen-free diamond wires predicted a narrow diameter-dependent stability interval, but omitted the catalyst and growth-environment surface chemistry. [35] *A* micrometer wire can have a small sidewall correction while still advancing through nanometer-scale islands.

### 3.3 Gas chemical potential

For a carbon reservoir equilibrated with methane and hydrogen, $\mu_C^{gas} = \mu_{CH4} - 2\mu_{H2}$. Relative to graphite, the ideal-gas expression is

$$\mu_C^{gas} - G_{graphite} = \Delta G_f{}^\circ(CH_4) + RT \ln\left[\frac{p_{CH_4}/p^\circ}{(p_{H_2}/p^\circ)^2}\right] \quad (3)$$

Here $\Delta G_f{}^\circ(CH_4)$ is the standard formation Gibbs energy from graphite and hydrogen, $R$ is the molar gas constant, p° is 1 bar and each pressure ratio is dimensionless. This equation is written per mole of carbon. The corresponding diamond supersaturation is smaller by $G_D - G_{graphite}$ at the same temperature and pressure. Standard-state data and numerical examples are given in SI Section S1.

At 1200 K with 1 mol% $CH_4$ and 99 mol% $H_2$, the calculated carbon potential relative to graphite is +18.8, −4.2 and −27.2 kJ mol$^{-1}$ at 0.1, 1 and 10 bar. Diamond lies about 6.8 kJ mol$^{-1}$ above graphite in the selected thermodynamic data, giving corresponding diamond driving forces of +12.0, −11.0 and −34.0 kJ mol$^{-1}$. [36, 34] Increasing methane to 5 mol% at 1 bar changes the diamond value to +5.9 kJ mol$^{-1}$. Positive values favor bulk precipitation from this idealized reservoir; a finite nucleus or wire also incurs surface and interface costs. Graphite remains more strongly favored in bulk.

At fixed methane and hydrogen fractions, increasing total ideal-gas pressure lowers this carbon potential by $RT \ln(P_2/P_1)$. This gas-composition effect is distinct from pressure favoring a denser condensed phase. At high pressure, fugacities and pressure-dependent solid free energies are required. In a flowing or activated feed, the catalyst need not equilibrate with the inlet gas; decomposition, transport and incorporation determine its actual carbon activity.

## 3.4 Structural mismatch

For an illustrative (111)product ‖ (111)starter contact with parallel [$\bar{1}$$\bar{1}$0] directions, the one-to-one mismatch is ($a_{\text{product}} - a_{\text{starter}}$)/$a_{\text{starter}}$, because the corresponding in-plane repeat is a/√2. Representative lattice parameters give −18.2% for diamond on 3C-SiC, −20.3% for cBN on BP and +1.4% for cBN on diamond (Table 2). [33, 37, 38, 19] Cubic symmetry alone therefore gives neither SiC nor BP a small one-to-one mismatch with its proposed product.

Table 2. Room-temperature one-to-one lattice mismatch for proposed cubic contacts. Values are geometric comparisons, not computed interface energies.

| **New solid / starter** | $a_{\text{product}}$ / $a_{\text{starter}}$ **(Å)** | **Mismatch (%)** |
|---|---|---|
| Diamond / 3C-SiC | 3.567 / 4.3596 | −18.2 |
| cBN / BP | 3.6165 / 4.5356 | −20.3 |
| cBN / diamond | 3.6165 / 3.567 | +1.4 |

*A* coincidence arrangement can reduce a geometric mismatch over several lattice periods, but it does not remove dislocations, chemical reconstruction or interfacial strain. Zhu and colleagues reported locally oriented diamond/SiC interfaces with defects and tilt. [39] Chen and colleagues resolved misfit accommodation at diamond/cBN interfaces. [40] Orientation and termination must therefore be specified before assigning an epitaxial advantage to a starter.

## 4 Kinetic considerations

Continuation requires precursor activation, delivery to the contact and completion of successive target layers before competing deposition or seed dissolution interrupts growth. Atoms may pass through a liquid, along a solid catalyst surface or through defects and buried interfaces. The relevant rates cannot be inferred from catalyst phase alone. [9, 41, 42] At the same incorporated-atom rate, a solid wire extends more slowly than a thin-walled nanotube of the same outer diameter because each unit of length contains more material.

Brus illustrated how a metastable silicon nanocrystal can persist because transformation from the β-tin structure to the diamond-cubic structure requires an activation barrier, $\Delta E$. [43] In that model, changes in particle shape and surface area contribute to the barrier. Equations 1 and 2 here give the bulk precipitation driving force and the incremental sidewall cost of elongation;

neither specifies the atomic pathway or its activation barrier. Attachment and competing-phase nucleation must therefore be treated as kinetic processes even when precipitation is thermodynamically favored.

The catalyst stores a finite inventory. Maliakkal and colleagues observed that changing Ga supply predominantly altered the wait for a new GaAs layer, whereas changing As supply predominantly altered layer-completion time. [44] Models by Glas and Dubrovskii show how a spreading layer can arrest when it exhausts available solute. [45] For geometrically similar contacts, stored material scales with catalyst volume, while one crystal layer scales with contact area. Shrinking the catalyst can therefore make vapor replenishment necessary during a single layer. SI Section S2 gives an illustrative inventory calculation and the corresponding component balances.

Changing temperature or feed can separate initiation from continuation. Wen and colleagues reported Si–Ge heterojunction sequences involving liquid and solid catalysts; Kodambaka and colleagues showed that catalyst thermal history can select liquid or solid Ge growth under comparable final conditions. [46, 47] Gamalski and colleagues observed solute-induced liquefaction of Au–Ge below the bulk eutectic and subsequent solidification as Ge precipitated. [48] *A* particle's state during growth must therefore be measured or constrained independently of its state after cooling.

Sheng and colleagues reported catalytic Joule-heating synthesis of nanowires within seconds. [49] Timed heating or feed interruption is useful if the desired and competing processes respond differently. *A* pulse must last long enough for target-layer completion while limiting competing nucleation, catalyst spreading or seed loss. Extension beyond the initial available inventory requires replenishment. Interrupted holds and cooling-rate comparisons determine whether incorporation occurs during the hold or during cooling. Comparisons should vary peak temperature, pulse duration and cooling separately; equal average temperature does not impose equal reaction histories. [50] Hydrogen-rich intervals need their own seed-retention control because metal-assisted diamond etching can accelerate under hydrogen. [51]

# 5 Candidate routes to diamond wires

## 5.1 Catalyst on an existing diamond seed

I would first revisit the seed–metal geometry of Derjaguin and colleagues. They reported whiskers with Ni, Fe and Mn droplets and specified an unidentified carbon-containing gas. Diffraction from detached fragments supported diamond, but did not establish continuous diamond structure through an intact metal-capped whisker or resolve substrate consumption. [6, 7] *A* modern test should follow a characterized catalyst footprint on diamond, trace carbon from a distinct vapor source and examine the extension separately from the seed.

*A* local catalyst deposit on a diamond facet provides the simplest direct test of the continuation hypothesis. I use an irregular 14-facet particle with eight {111} and six {100} faces because it places two common diamond surface families on the same crystal and permits comparison of face-center, edge and multiface contacts (Figure 1). Unequal facet areas represent the irregular particles of interest; fourteen facets are not a requirement for growth. Catalyst size and footprint should be compared after heating, when spreading and dewetting may have changed the deposited pattern.

Commercial HPHT diamond grit offers an economical starting material for comparing many catalyst deposits and growth conditions. Huanghe Whirlwind, for example, offers HFD-*D*/C grit described as complete cubo-octahedral crystals. [52] The ideal cuboctahedron has eight triangular {111} faces and six square {100} faces. *A* deposit covering a {111} face would meet three edges, compared with six on the truncated-octahedral {111} face illustrated here; the {100} face has four edges in both geometries. Comparing these particles could test how the facet perimeter affects contact-line pinning and retention during growth.

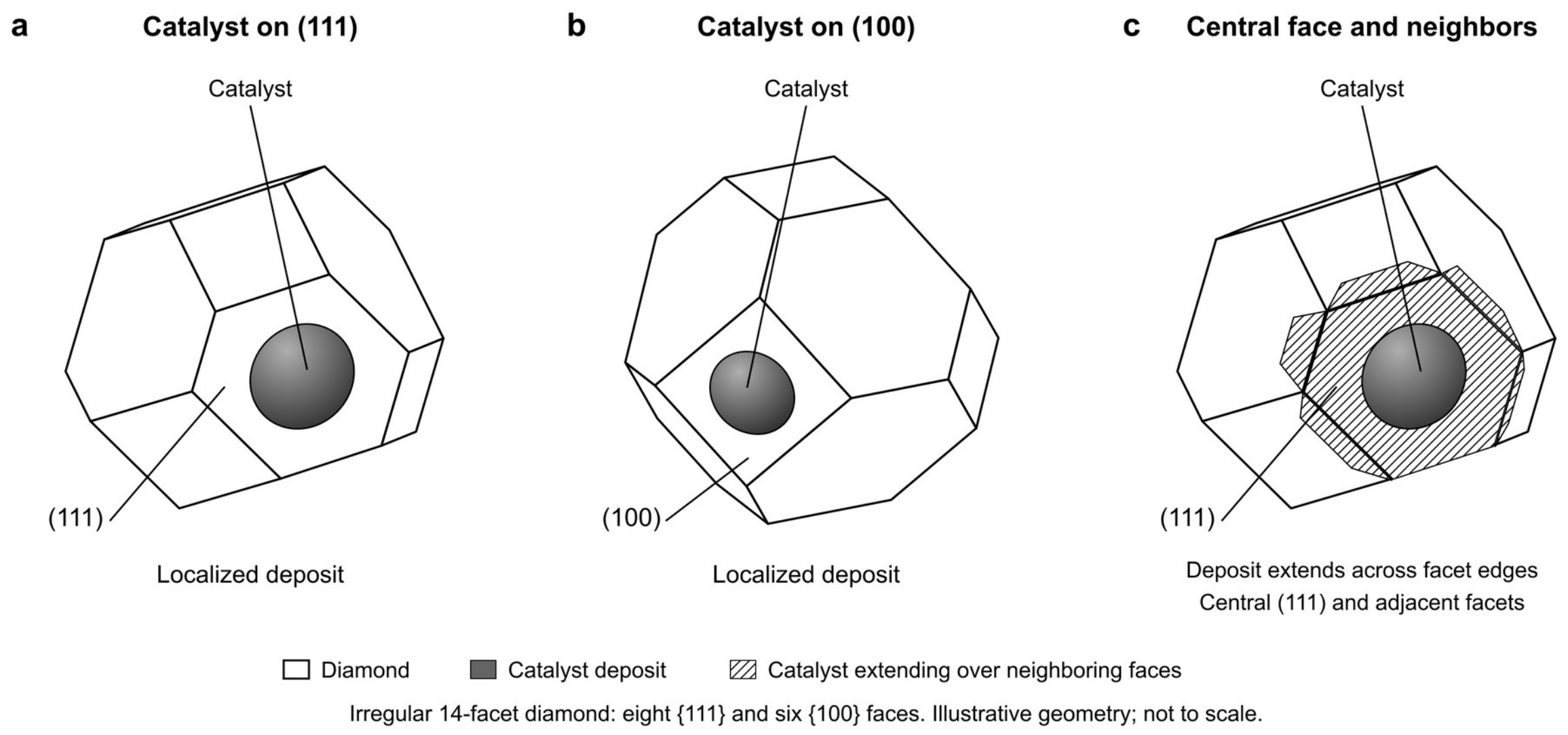


Figure 1. Catalyst placement on an irregular diamond particle with eight {111} and six {100} faces. a *A* rounded deposit contacts a (111) face. b *A* rounded deposit contacts a (100) face. c *A* connected deposit covers a central (111) face and its six edge-sharing neighbors. The shape allows facet identity and contact geometry to be varied on one crystal. Deposits and particles are schematic; the drawings do not predict equilibrium wetting.

Initial trials can coat several facets. Directional deposition, temporary embedding or local transfer can subsequently restrict the contact if location affects extension. Reports of facet-selective Ag and Fe-based deposition show that preparation and annealing can redistribute metal or etch the diamond. [53, 54, 55] Full-face deposits might be retained by facet edges, although a liquid need not wet every polygon corner. [56] SI Figure S1 illustrates the proposed full-facet contact geometries.

*A* metal bridge can also contact a diamond particle on one side and a larger diamond plate on the other. Different particle and plate facets permit a test of orientation-dependent dissolution and deposition (SI Figures S2 and S3). Different dissolution rates alone cannot sustain transfer between otherwise equivalent isothermal diamond reservoirs. The source chemical potential must exceed the incremental cost of incorporation at the receiving contact, including new sidewalls; facet kinetics then influence where transfer occurs. Vapor labeling is needed to distinguish eventual gas-fed extension from growth supplied only by plate recession. Transfer supplied only by a solid through liquid metal is solid–liquid–solid growth; vapor incorporation is required for VLS.

Preformed diamond wires and exposed diamond ends in partially etched anodic-alumina templates offer another seeded contact. Reports include polycrystalline cylinders, nanocrystalline pillars and structures assembled from diamond particles; the hollow tubes in Masuda and colleagues' study were diamond-like carbon. [57, 58, 59] Single-crystal diamond wires made by etching provide a less ambiguous initial lattice. [22] Metal deposited at an identified diamond end could test continued elongation, while an annular end would test whether a hollow structure continues or fills.

### 5.2 Switching a SiC starter to diamond

*A* SiC starter tests initiation on a different crystal. The experiment is to grow a short capped SiC wire, retain its catalyst and change silicon and carbon supply independently (Figure 2). Thirumalai and colleagues reported Ni-assisted SiC wire growth with separate precursor delivery. [60] Diamond growth on SiC provides an interface precedent, although the large one-to-one mismatch and reported defects preclude assuming coherent continuation. [39]

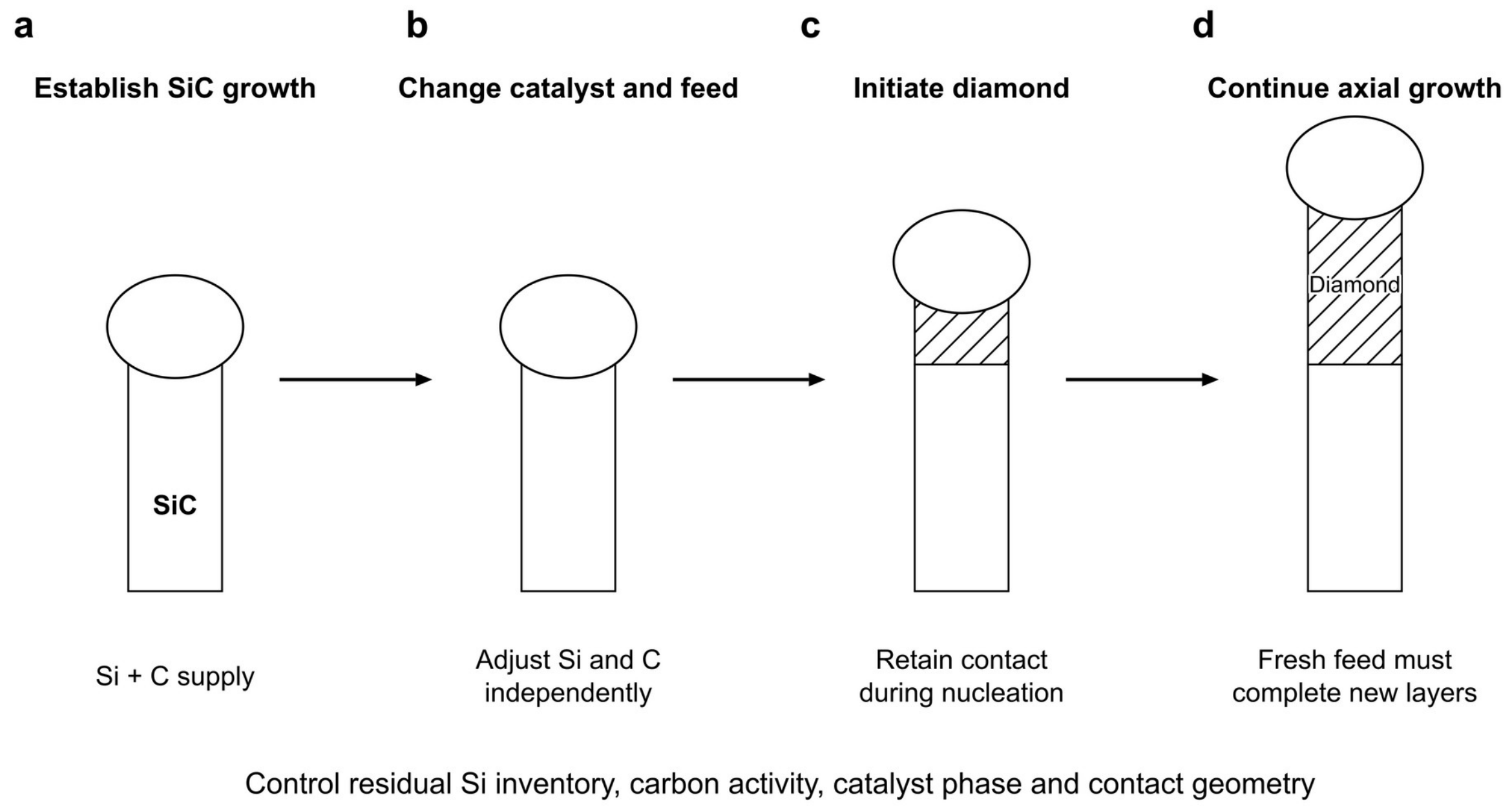


Figure 2. Proposed switch from a SiC starter to diamond. a Establish a capped SiC wire. b Reduce silicon supply while controlling carbon supply. c Initiate diamond at the retained contact. d Test repeated extension from fresh vapor. The transition may be abrupt, defective or multiphase. The catalyst and SiC are unfilled; diamond is diagonally hatched. Shapes and dimensions are schematic.

Stopping the silicon precursor does not immediately remove silicon from the catalyst. Stored silicon and starter dissolution can sustain SiC growth after the switch. At fixed silicon chemical

potential, raising carbon activity favors both SiC and diamond precipitation; reducing silicon supply is the additional operation that can suppress continued SiC. I would compare a complete interruption with a small residual silicon feed, because residual silicon could preserve the starter or affect carbon incorporation. Our liquid-metal diamond study showed a strong dependence on silicon content. [26]

The decisive observation is a structurally identified diamond segment extending from the retained tip and incorporating the new carbon feed. *A* shell around an eroding SiC wire, a finite conversion layer or graphite at a separated catalyst does not establish this transition. Matched abrupt and gradual switches can reveal whether the catalyst inventory or contact reconstruction limits initiation.

Other carbides, GaN and $Ga_2O_3$ introduce different starter-dissolution and feed-switching constraints. [61, 62, 63, 64, 65, 66] Carbon black or platelet graphite can instead supply local carbon without supplying a diamond lattice. [67, 68] Fluorinated interwall-bonded carbon structures are more speculative starters, motivated by diamane and collapsed-tube calculations; they require direct verification of interwall C–C bonding and survival of the catalyst-contact preparation. [69, 70] These alternatives are grouped by their distinct initiation requirement in SI Section S3.

## 6 Candidate routes to cubic boron nitride wires

The first cBN test should likewise use an identified cBN seed, so failure of continued incorporation can be separated from failure to nucleate the cubic phase. *A* catalyst deposit could contact a cBN particle or a verified cBN region on diamond (Figure 3). Reported diamond/cBN interfaces and cBN deposition on diamond make the latter structurally plausible, but neither proves vapor-fed wire continuation. [40, 71] The cBN–catalyst contact must survive conditioning without a blocking hBN layer. Where the catalyst also contacts diamond, carbon uptake can change the catalyst composition and introduce competing carbon or B–C phases. [72]

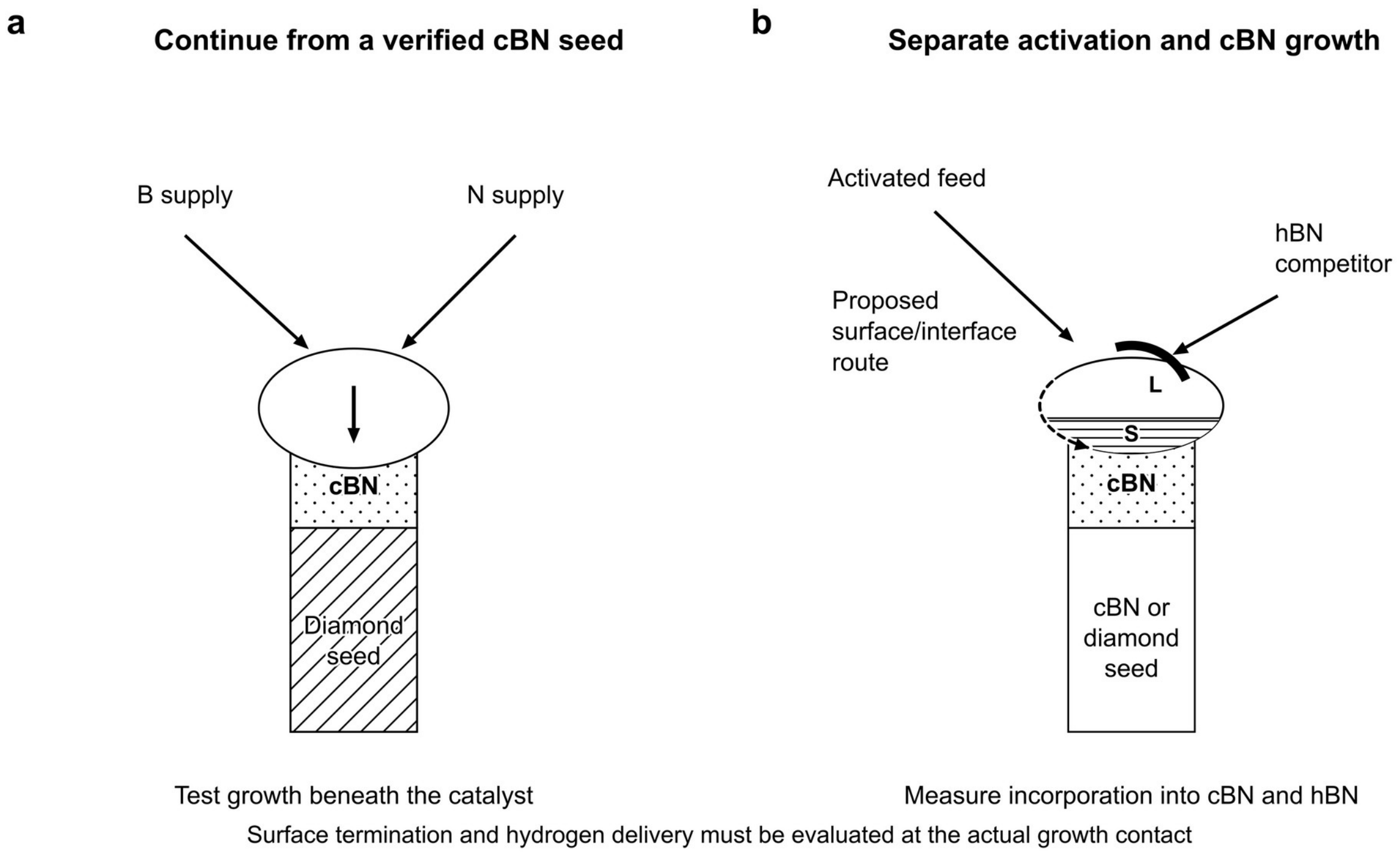


Figure 3. Proposed cBN continuation from an established contact. a *A* verified cBN nucleus on a terminal diamond facet receives catalyst and fresh B and N; an existing cBN tip is another starting point. b *A* liquid supply region connects to a solid contact region, with the dashed arrow indicating a possible surface or interface transport path. hBN can compete at the supply surface. Horizontal hatching denotes solid catalyst, stippling cBN, diagonal hatching diamond and black hBN. *L* and S label liquid and solid regions. Both panels are proposed geometries, not demonstrated growth.

Boron and nitrogen need not enter by the same route. *A* catalyst that dissolves boron but little nitrogen could receive reactive nitrogen at its exposed surface or contact perimeter; another catalyst could transport both through its interior. Nickel-boride-assisted BN nanotube growth provides a vapor-fed experimental comparison, while seeded HPHT cBN growth supplies a positive control for the target crystal in a solvent. [10, 28] *A* nitrogen plasma is an option for testing activation limitations.

Cheng, Bets and Yakobson calculated BN nucleation at steps on diamond(001). [73] Their model favored cBN over an intermediate range of hydrogen availability. Without hydrogen, the growing BN became disordered. At higher hydrogen chemical potential, stronger hydrogen passivation inhibited bonding between the BN nucleus and diamond and favored $sp^2$ BN. These calculations concern nucleation directly on diamond; the corresponding conditions at a metal–BN interface remain to be determined.

I would first measure seed recession, catalyst phase and boron uptake under the intended conditioning steps. B and N supply would then be varied separately, including a nitrogen-only exposure to identify seed conversion or finite nitridation of material already stored in the particle. Cubic extension over repeated replenishment cycles, with incorporation of new B and N, is the desired result. Boron-rich precipitation, hBN encapsulation and nitridation of the catalyst indicate different limits and call for different changes in feed.

*A* BP starter offers a tetrahedral but strongly mismatched contact; replacing phosphorus supply with nitrogen also requires removal of phosphorus stored in the catalyst. [74] Boron, boron carbide and $CaB_6$/$SrB_6$ starters instead offer boron inventory and a pre-existing contact, without an established cBN orientation relationship. [75, 76, 77, 78, 79, 80] Ca or Sr released by dissolution remains chemically relevant after its external supply stops. Functionalized double-wall BN nanotubes are a further candidate only if interwall B–N bonding produces an ordered tetrahedral region; B–F bonds alone do not establish such a seed. [81, 82, 83]

## 7 Calculations and experiments to test the proposals

### 7.1 Initial experiments and comparisons

Table 3 specifies initial comparisons for the three experiments. Published conditions provide reference points; the temperatures, particle sizes, time points and feed variations proposed here are not optimized. In particular, 1025 °C is a reported condition for one liquid alloy, not a required starting temperature or an established optimum for wire growth. Temperature series should be chosen for each catalyst and contact, considering its phase, precursor activation, seed dissolution and competing deposition, and extended beyond the illustrative values in Table 3 when those observations warrant it. Growth tests measure initiation frequency, length versus hold time, phase of the extension and catalyst position across the sampled contacts.

Table 3. Ranked initial experiments. Screening choices are proposed here and are not established wire-growth conditions. Additional definitions and controls are given in SI Section S4.

| **Priority and starting contact** | **Proposed initial conditions** | **Comparison and required observation** |
|---|---|---|
| 1 Diamond seed/metal | Indexed {100}/{111} seeds; separate Ni and $Ga_{77.75}Ni_{11}Fe_{11}Si_{0.25}$ contacts (atomic %), 0.1/1/10 µm footprints. An illustrative temperature series is 975/1025/1075 °C at 1 atm, $CH_4$:$H_2$ = 5:100; hold 5/15/30/150 min. [26] supplies one alloy-specific reference, not a temperature optimum. Adapt the range to each contact. | Bare seed, unseeded catalyst, no-carbon feed and $^{13}CH_4$ comparisons. Require repeated diamond extension with new vapor labels, distinct from seed recession and cooling deposits. |
| 2 SiC starter switch | Retain Ni-containing caps on 0.1–0.3 µm SiC wires and isolate them from the solid SiC source. [84] An illustrative restart series is 975/1025/1075 °C, 1 atm, $CH_4$:$H_2$ = 5:100; hold 1/5/15/30 min. These restart conditions are unvalidated; choose and extend the temperature range according to catalyst phase and starter survival. | Compare continued SiC supply, carbon-only restart, feed-free and cap-free starters. Require local diamond at the former tip contact, increasing length and fresh carbon incorporation; track retained Si. |
| 3 cBN seed/boride | $Ni_2B$ contacts, 0.5–2 µm, on 10–100 µm cBN seeds; 1000/1100 °C; 1 bar $N_2$ with calibrated 0.01/0.1 mol% borazine; hold 1/5/15/30 min. Temperature/chemistry follow BNNT precedent [10]; dose, sizes and time grid are proposed. | Uncapped seed, unseeded boride and precursor-free controls; separate B and N labeling. Require cubic extension through several fresh-feed intervals. Boride inventory, hBN and seed loss must be resolved. |

## 7.2 Provenance and structural identification

Vapor-derived material must be identified within the new extension. For diamond, alternating $^{12}CH_4$ and $^{13}CH_4$ can distinguish feed intervals from the original seed; for BN, $^{10}B$/$^{11}B$ precursors and a separate nitrogen isotope test distinguish the two sources. CNT isotope barcoding supplies a useful precedent. [85, 86, 87] Gas transit, catalyst storage and isotope exchange broaden the composition boundaries. Labeling pulses must match precursor purity and delivery, since an impurity introduced with a labeled gas can itself alter nucleation. [86]

NanoSIMS can map isotope ratios along sufficiently long segments. Its best lateral probe sizes are about 50 nm, but isotope precision and usable segment resolution must be established on a reference under the same conditions. [88] I would initially target labeled intervals several

hundred nanometers long. Atom-probe tomography offers finer three-dimensional analysis: Schirhagl and colleagues measured carbon isotopes and a nanometer-scale nitrogen profile in boron-doped diamond. [89] Application to cBN requires validation of field evaporation and mass-peak assignments. Raman isotope shifts can identify longer labeled regions, while local diffraction and conventional core-loss electron energy-loss spectroscopy (EELS) address structure and bonding.

Monochromated vibrational EELS in a scanning transmission electron microscope (STEM) provides a further isotope-sensitive measurement. Senga and colleagues distinguished $^{12}C$- and $^{13}C$-rich regions in graphene at subnanometer resolution, and Li and colleagues measured isotope-dependent phonons across an *h*-$^{10}BN$/*h*-$^{11}BN$ interface. [90, 91] Applying this method to diamond or cBN wires would require isotope reference spectra and assessment of phonon delocalization, strain and beam damage.

Isotope composition identifies provenance, not crystal structure. Correlative diffraction and bonding measurements must sample the extension away from the seed, with cross-sections through its catalyst contact where possible. Extended hexagonal stacking must be distinguished from isolated faults or mixed cubic/hexagonal sequences. [92, 93] Seed recession and fixed position markers distinguish added material from remnants exposed by etching. *A* conservative inventory bound on seed, starter, local source and catalyst provides an independent check when the new volume exceeds all plausible stored contributions.

### 7.3 Interpreting interrupted growth

Interrupted runs separate incorporation during the hold from precipitation on cooling. Where in situ observations are feasible, the probe can itself change the reaction: Cheek and colleagues reported electron-beam-driven Ge growth in liquid-cell microscopy. [94] Catalyst-free, unseeded, feed-free and probe-exposure controls address distinct alternatives and should be selected for the ambiguity in the particular experiment.

Continued SiC indicates persistent silicon activity; rapid starter recession indicates loss before target initiation; an isolated target crystallite with no later extension suggests inadequate replenishment or layer completion; an encapsulated catalyst indicates competing surface deposition. Predicting which catalyst is best would additionally require alloy activities, interfacial free energies and attachment barriers that the present estimates do not supply.

Density functional theory (DFT) calculations could compare interface energies and barriers for atomic attachment or competing-phase nucleation. Simulations using machine-learned interatomic potentials could follow transport and incorporation over longer times. Hedman and colleagues used DFT-trained potentials to study CNT growth and defect repair on Fe nanoparticles; Hedman also studied CNT growth and surface carbon transport on solid Re nanoparticles. [95, 96] For diamond or cBN, such potentials would need validation for the target and competing phases, catalyst interfaces and relevant compounds. CALPHAD (calculation of phase diagrams) assessments could provide catalyst phase equilibria and component activities.

## 8 Outlook

I would begin with diamond seed–metal contacts, compare them with SiC starter switches, and test cBN seed retention and replenishment in parallel. The first convincing outcome would be a reproducible target-phase segment whose length increases through several intervals of fresh vapor supply while its growth contact is retained. The diameter should remain a measured variable, including micrometer and larger wires for which sidewall costs are smaller.

## Methods

### Illustrative calculations

Sidewall costs were evaluated for fixed-cross-section elongation; ideal-gas carbon potentials used tabulated formation Gibbs energies; lattice mismatches used the stated room-temperature parameters and orientations. SI Section S1 gives the equations, inputs and assumptions. No fitted growth model, atomistic simulation or new CALPHAD assessment was performed.

### *Use of artificial intelligence*

During preparation and revision of this manuscript, I used OpenAI ChatGPT/Codex to assist with literature searches, organization and analysis; assessment of my proposed growth approaches; drafting and revision of text; checking of equations, notation, internal consistency and references; implementation and checking of illustrative numerical calculations; and preparation of programmatic schematic figures. I directed the scientific scope and take responsibility for the manuscript’s scientific arguments, equations, citations and graphics.

## Supporting Information

Numerical inputs and thermodynamic estimates; catalyst inventory and supply balances; additional starter and contact comparisons; isotope measurements and property tests; three figures and three tables. *A* reproducible calculation script is supplied separately.

# Supporting Information

## Toward catalyst-mediated growth of diamond and cubic boron nitride wires

Rodney S. Ruoff[1,2,3,4,*]

[1] Center for Multidimensional Carbon Materials (CMCM), Institute for Basic Science (IBS), Ulsan 44919, Republic of Korea
[2] Department of Chemistry, UNIST, Ulsan 44919, Republic of Korea
[3] Department of Materials Science and Engineering, UNIST, Ulsan 44919, Republic of Korea
[4] School of Energy and Chemical Engineering, UNIST, Ulsan 44919, Republic of Korea
[*] Corresponding author: ruofflab@gmail.com

This Supporting Information provides numerical inputs, illustrative estimates, additional contact geometries and measurement details for the Perspective.

### S1 Numerical thermodynamic comparisons

The estimates below evaluate terms in the main-text argument using published thermochemical data and room-temperature lattice parameters. They do not predict a catalyst composition or a wire-growth window. Standard-state choices and input values are given below and in the accompanying calculation script so that each value can be reproduced.

The representative cubic lattice parameters are 3.567 Å for diamond, 3.6165 Å for the $^{10}$B-enriched cBN crystal measured by Chen and colleagues, 4.3596 Å for 3C-SiC and 4.5356 Å for BP. [S1, S2, S3, S4] The diamond datum is a near-ambient, room-temperature database value; these are not growth-temperature measurements of the proposed contacts. The conventional cubic cell contains eight carbon atoms in diamond and four BN pairs in cBN, giving $\Omega_D = a_D^3/8$ and $\Omega_{cBN} = a_{cBN}^3/4$. Energies are normalized per C atom or per BN pair, not per total atom in both materials. Conversion uses 1 eV per particle = 96.485 kJ mol$^{-1}$.

Table S1. Sidewall cost for a circular wire. The surface-energy interval is a selected sensitivity range, not a measured catalyst–wire interface energy.

| Diameter (nm) | Diamond (eV/C atom) | cBN (eV/BN pair) |
|---|---|---|
| 10 | 0.0142 (0.00142–0.0708) | 0.0295 (0.00295–0.148) |
| 100 | 0.00142 (0.000142–0.00708) | 0.00295 (0.000295–0.0148) |
| 1000 | 0.000142 (1.42e-05–0.000708) | 0.000295 (2.95e-05–0.00148) |
| 10000 | 1.42e-05 (1.42e-06–7.08e-05) | 2.95e-05 (2.95e-06–0.000148) |

Central values use $\gamma_{side}$ = 1 J m$^{-2}$; parentheses give the selected 0.1–5 J m$^{-2}$ range.

For an increment dL, the new sidewall free energy is $\Sigma_j \gamma_j \ell_j$ dL and the incorporated amount is $A$ dL/$\Omega$. Their ratio gives main-text Eq. 2. This is the free-energy derivative for fixed-cross-section elongation. *A* changing cap area, contact angle, strain or catalyst composition adds separate terms. The tabulated $\gamma$ range cannot be used to rank diamond against a nanotube unless the competing surfaces and equal-material geometry are also evaluated.

Table S2. Ideal-gas carbon chemical potential relative to graphite and diamond. Values refer to the stated gas reservoir, not a measured catalyst activity.

| ***T* (K)** | **P (bar)** | **$CH_4$ (mol%)** | **Relative to graphite (kJ mol$^{-1}$)** | **Relative to diamond (kJ mol$^{-1}$)** |
|---|---|---|---|---|
| 1000 | 0.1 | 1 | +0.5 | -5.4 |
| 1000 | 1 | 1 | -18.6 | -24.5 |
| 1000 | 10 | 1 | -37.8 | -43.7 |
| 1200 | 0.1 | 1 | +18.8 | +12.0 |
| 1200 | 1 | 1 | -4.2 | -11.0 |
| 1200 | 10 | 1 | -27.2 | -34.0 |
| 1400 | 0.1 | 1 | +37.2 | +29.5 |
| 1400 | 1 | 1 | +10.4 | +2.7 |
| 1400 | 10 | 1 | -16.4 | -24.1 |
| 1200 | 1 | 5 | +12.7 | +5.9 |

The remaining gas fraction is $H_2$. Positive values favor bulk deposition from the stated reservoir.

NIST-JANAF gives standard methane formation Gibbs energies of 19.492, 41.524 and 63.761 kJ mol$^{-1}$ at 1000, 1200 and 1400 K, respectively, relative to graphite and $H_2$ at a standard pressure of 1 bar. [S5] The graphite and hydrogen contributions are already included in these formation values. Robie and Hemingway give diamond–graphite differences of 5.9, 6.8 and 7.7 kJ mol$^{-1}$ at those temperatures. [S6] The tabulated gas cases use ideal partial pressures and compare solids at 1 bar; estimated condensed-phase pressure corrections over 0.1–10 bar are below 0.005 kJ mol$^{-1}$

using constant molar volumes. No high-pressure extrapolation is made. The two compilations support illustrative estimates; a precise phase-boundary calculation should use an internally consistent thermodynamic assessment. Printed digits and the selected $\gamma$ range are not propagated measurement uncertainties.

Table 2 uses the starter in the denominator and room-temperature lattice parameters. *A* geometric five-to-four comparison gives $(5a_{\mathrm{D}} - 4a_{\mathrm{SiC}})/(4a_{\mathrm{SiC}}) = +2.27\%$ and $(5a_{\mathrm{cBN}} - 4a_{\mathrm{BP}})/(4a_{\mathrm{BP}}) = -0.33\%$. These are illustrative numerical coincidences, not demonstrated orientation relationships or calculations of interface stability. The one-to-one comparison remains the direct measure of adjacent-period mismatch.

The mismatch comparison uses parallel cubic axes and corresponding in-plane translations of the same low-index plane. The common factor relating each translation to its cubic lattice parameter cancels in the one-to-one ratio. The result does not establish the favored interface orientation. *A* multi-period coincidence is a different comparison and must state both integer periods and its denominator. Thermal expansion, polarity, termination and relaxation are omitted.

## S2 Catalyst inventory and time-dependent supply

For component *i*, the catalyst balance is input from vapor and local solids minus crystal incorporation and all other losses. Stored material includes dissolved atoms, surface material and compounds; only the fraction able to reach the growth front during the relevant interval is available for layer completion.

$$\frac{\mathrm{d}N_i}{\mathrm{d}t} = F_i^{\mathrm{gas}} + F_i^{\mathrm{wire}} + F_i^{\mathrm{surface}} - F_i^{\mathrm{loss}} - \sum_{\alpha} \nu_{i\alpha} \frac{\mathrm{d}n_\alpha}{\mathrm{d}t} \tag{S1}$$

Here $N_i$ is the number of atoms of component *i* in the catalyst, *t* is time and each F is an atom rate. The superscripts gas, wire and surface identify supply paths; loss includes desorption and other noncrystalline losses. For each precipitating phase $\alpha$, $n_\alpha$ is its number of formula units and $\nu_{i\alpha}$ is the number of atoms of component *i* per formula unit. Input and loss terms exclude atoms already counted in the phase-incorporation term. *A* boride or carbide can change both inventory and chemical activity, so its entire elemental content must not be treated as freely available solute. For BN, the smaller stoichiometric supply of B and N limits the number of complete layers.

For illustration, take a hemispherical catalyst of radius $r$ and atomic density $n_{cat}$, a circular wire of the same radius and atomic density $n_{wire}$, and a layer height $h$. If $x_{av}$ is the fraction of catalyst atoms available as carbon during layer completion, the initial inventory in diamond-layer equivalents is $(2/3)x_{av}(n_{cat}/n_{wire})(r/h)$. With deliberately selected $x_{av} = 0.01$, $n_{cat}/n_{wire} = 0.5$ and $h = 0.20$ nm, catalyst diameters of 10, 100 and 1000 nm store approximately 0.083, 0.83 and 8.3 layer equivalents. These assumptions do not specify any actual alloy solubility. They show why a small cap may require continuing supply even during one layer; for cBN, the B and N balances must be evaluated separately.

*A* diffusion time estimated as $L^2/D$ only compares a chosen transport length $L$ with a specified diffusivity $D$; it does not include precursor activation, segregation or attachment. *A* useful kinetic calculation couples this redistribution to the measured refill rate and the time needed to complete one layer. Finite-inventory layer arrest and composition-dependent catalyst phase changes have precedents in other wire systems. [S7, S8, S9, S10, S11]

*A* pulse comparison should match final feed composition and integrated precursor delivery where practical, measure the temperature history, and vary peak temperature, hold duration and cooling independently. Equal mean temperature is insufficient for reactions with nonlinear temperature dependence. *A* labeled precipitate appearing only on cooling is distinct from extension during the growth hold. The maximum wire length supplied by a finite solid source is its available target-element inventory divided by the amount required per unit length; for BN the smaller B- or N-based bound applies.

## S3 Contacts and alternative starters

The main experiments separate a seed that supplies a target lattice from a starter that must first support target nucleation and from feedstock that supplies atoms. Table S3 lists additional starting structures and the requirement each introduces.

Table S3. Additional starting structures and the requirement each introduces.

| **Starting structure** | **What it supplies** | **Additional requirement** |
|---|---|---|
| Diamond ends in AAO or etched diamond wires | An exposed target-phase end. [S12, S13, S14, S15] | Verify diamond at the exposed end; add catalyst and distinguish further growth from etching or filling of an annular contact. |

| Starting structure | What it supplies | Additional requirement |
| --- | --- | --- |
| GaN, $Ga_2O_3$ and other carbides | *A* wire/catalyst contact with different reaction chemistry. [S16, S17, S18, S19, S20, S21] | Retain the contact while removing the original component supply; quantify damage, dissolved starter elements and competing phases. |
| Carbon black or platelet carbon | Local carbon feedstock; no diamond lattice. [S22, S23] | Nucleate diamond separately and show continued vapor incorporation after bounding source consumption. |
| Functionalized double-wall carbon or BN tubes | *A* possible tetrahedral patch, if interwall bonds form. [S24, S25, S26, S27, S28] | Expose exterior and bore-facing surfaces to functionalization; verify ordered interwall C–C or B–N bonding. C–F or B–F bonds alone do not establish linking. Retain the tetrahedral region during catalyst preparation. |
| Boron wires or sheets, boron carbide | B inventory and an existing shape. [S29, S30, S31, S32] | Remove or characterize surface layers; distinguish finite nitridation from replenished cBN growth. |
| $CaB_6$ and $SrB_6$ | *A* B-rich starter with metal-containing catalyst history. [S33, S34] | Track Ca/Sr released by dissolution and distinguish removal of feed from removal of the metal inventory. |
| BP | Tetrahedral zincblende structure with large one-to-one mismatch. [S35] | Replace P supply with N; quantify retained P and the interface actually formed. |

Directional deposition on an irregular diamond particle will generally reach inclined neighboring faces. *A* partial embedding layer, aperture or local transfer can restrict deposition when a facet comparison requires it. Metal placement must be measured after conditioning: Ag redistribution, oxidative microfaceting and metal-assisted etching can change both footprint and surface orientation. [S36, S37, S38] Equal deposited metal mass does not imply equal contact area.

The six-edged {111} footprints in Figure S1 refer to the illustrated truncated-octahedral geometry. Cuboctahedral diamond grit provides a comparison with triangular {111} faces and four-edged {100} faces. [S39] At matched facet orientation and comparable contact area, the three-edge and six-edge perimeters allow tests of contact-line retention during heating and growth. The deposited footprint and its evolution should be measured, since a liquid may detach from edges or leave corners unwetted.

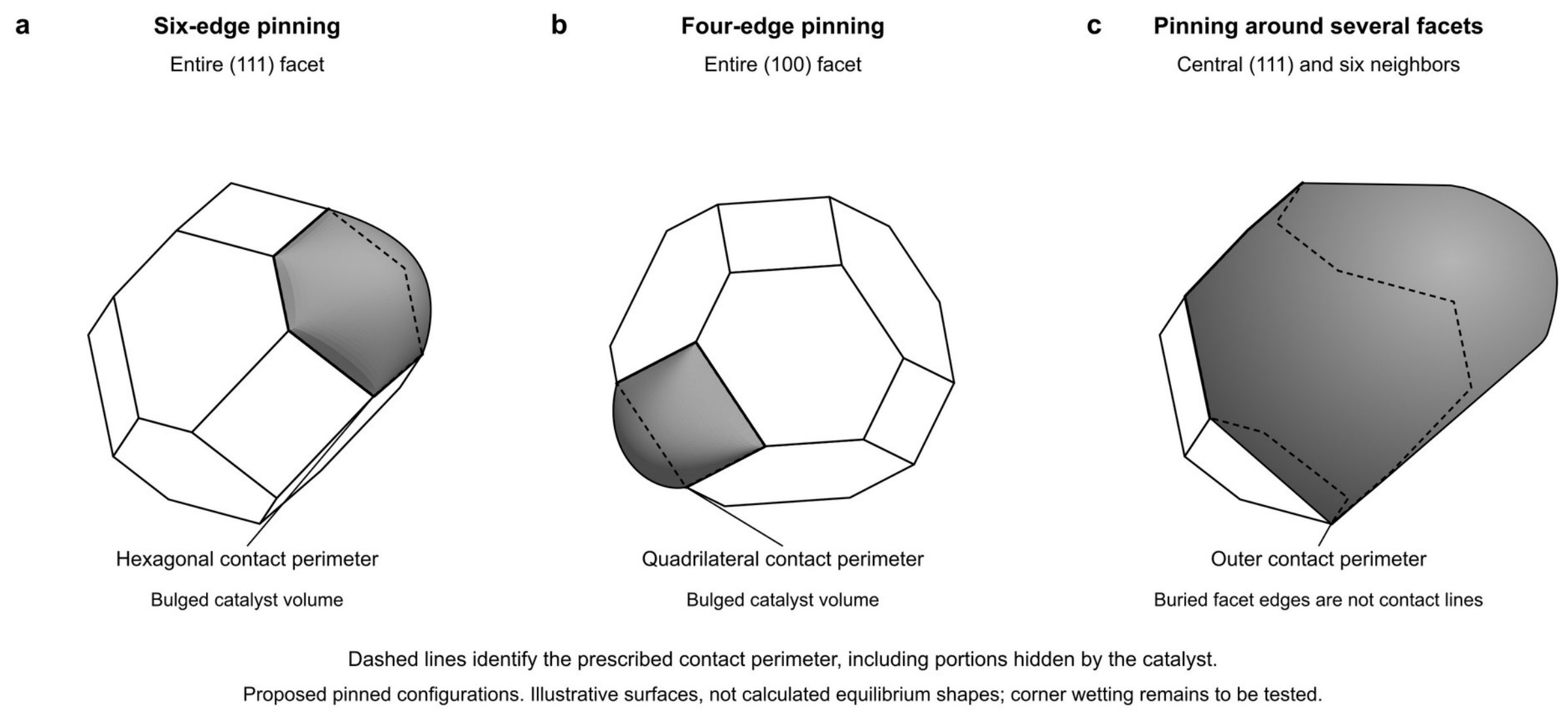

Figure S1. Proposed edge-pinned catalyst footprints: a, a six-edged {111} facet; b, a four-edged {100} facet; c, connected neighboring facets. Dashed lines show the outer contact line, including hidden portions; buried shared edges are not contact lines. These prescribed footprints test retention; they are not predicted equilibrium droplet shapes. Corners can remain dry, and growth can alter the facet beneath the contact. [S40]

*A* source plate and seed particle can contact a common metal region. Facet-dependent rates may determine where material dissolves or attaches, but equal-temperature bulk diamond surfaces share the same bulk chemical potential. *A* persistent transfer experiment therefore needs a specified driving force, such as curvature, stress, temperature difference or an imposed chemical reservoir. Seed uplift alone would not identify its source. Recession maps and isotope-distinct vapor separate plate-derived growth from vapor-derived extension.

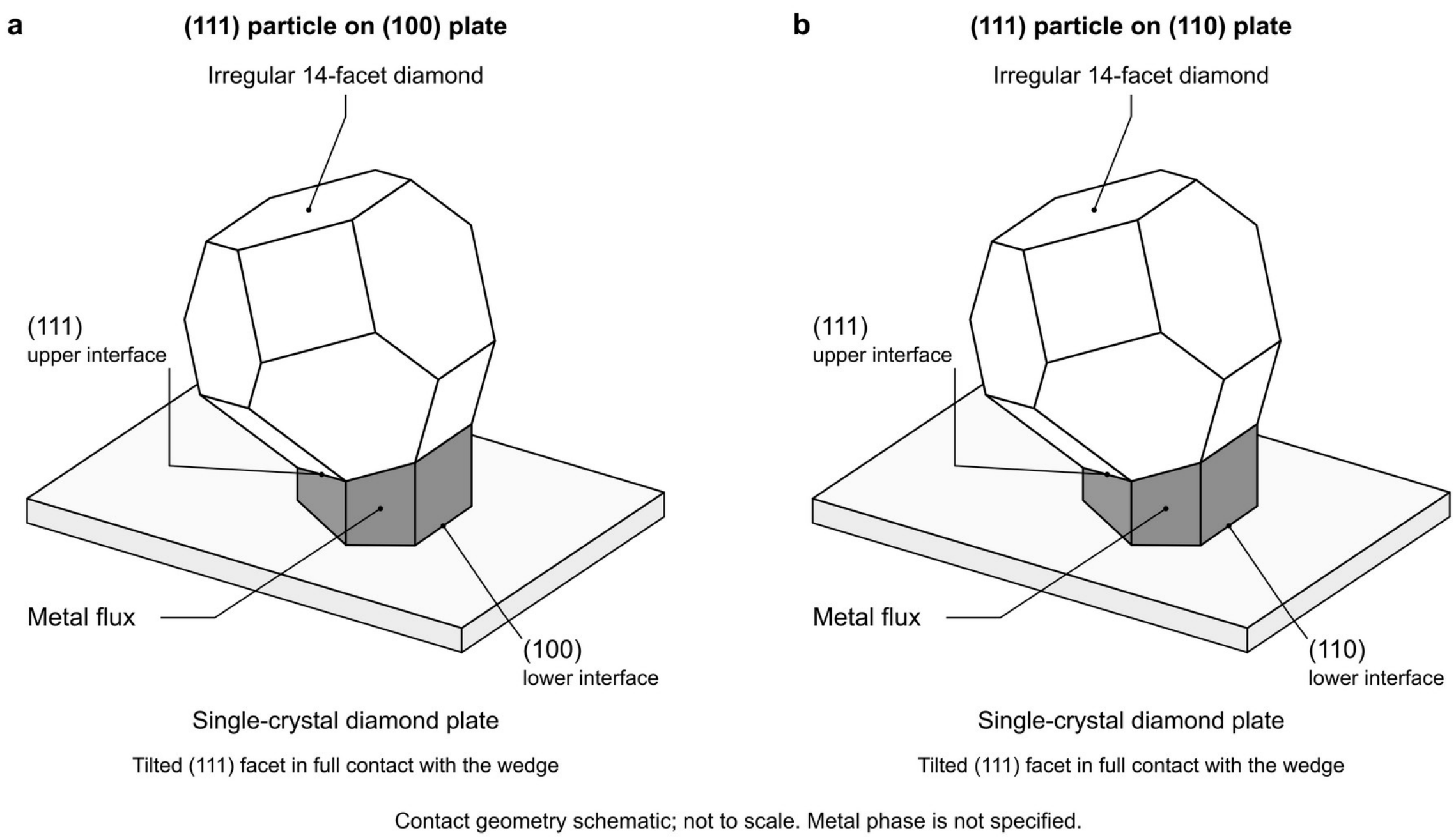

Figure S2. Proposed metal-wedge contacts between a faceted diamond particle and a diamond plate. The particle presents an inclined (111) face to the metal; the plate presents (100) in a and (110) in b. These geometries permit tests of contact retention and facet-dependent kinetics without assuming that different facet indices alone drive sustained transfer.

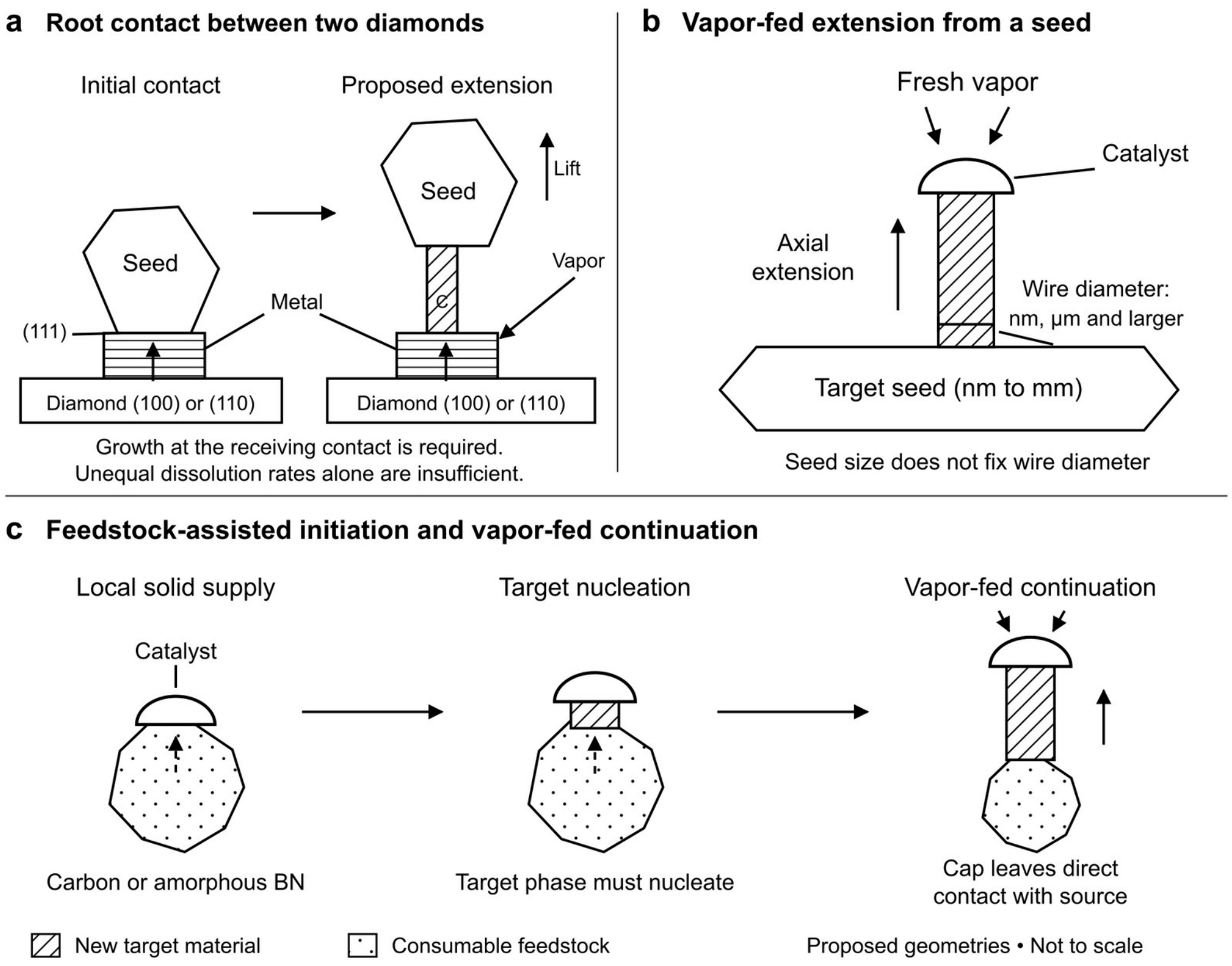

Figure S3. Seed and local-source configurations. a *A* polyhedral seed contacts metal on a diamond plate. b Diamond or cBN seed size and catalyst footprint vary independently. c Non-diamond carbon or non-cubic BN supplies material to the catalyst. Diagonal hatching denotes new diamond or cBN, horizontal hatching the bridge in a, and stippling feedstock in c. Catalysts in b and c are unfilled. The seed supplies a target lattice; the consumable source need not do so. Continued gas supply must be distinguished from finite growth by source consumption.

## S4 Experimental comparisons and isotope measurements

Gong and colleagues' liquid-metal experiment provides a reference at approximately 1025 °C, 1 atm and a $CH_4$:$H_2$ molar flow ratio of 5:100. Their wall pyrometer read 1175 °C; that is not the reported local growth temperature. [S41] Neither value defines a preferred temperature for the proposed contacts. The 975/1025/1075 °C series in Table 3 is an illustrative first comparison, not a prescribed window or a requirement to begin at 1025 °C. Each catalyst and starter may require a substantially different range. Temperature, contact survival, precursor activation and competing deposition should be assessed together. The proposed 0.1, 1 and 10 µm contact sizes and 5, 15, 30 and 150 min interrupted runs are new comparisons; $CH_4$:$H_2$ = 1:100/5:100/10:100 gives a separate feed series. Pure Ni and the Ga-rich alloy must not be assumed to have the same phase. For SiC, physically separate a retained-cap starter from its solid SiC source before admitting the independently supplied carbon feed; merely stopping a carrier gas does not prove the source has stopped supplying silicon. [S42]

For every initial screen, record the conditioned catalyst composition and footprint, seed dimensions and phase, gas arrival time and measured sample temperature. Catalyst-free seeds test exposed-surface deposition. Unseeded catalyst contacts test target nucleation under the continuation conditions. Feed-free runs test local-source and stored-solute contributions under that gas condition. Absence of growth does not exclude seed participation after the precursor changes catalyst chemistry. Interrupted holds and varied cooling rates locate when precipitation occurs.

The cBN screen adapts the NiB/$Ni_2B$ and borazine-containing vapor chemistry used for BN nanotubes at 1000–1100 °C. [S43] The proposed 1 bar pressure, metered 0.01/0.1 mol% borazine, 0.5–2 µm contacts on 10–100 µm cBN seeds and 1/5/15/30 min holds are selected tests, not values established for cBN growth. Borazine couples initial B and N delivery; independently controlled $B_2H_6$/$NH_3$ feeds would be a later comparison after identifying a seed-surviving contact. Separate boron- and nitrogen-isotope runs distinguish precursor incorporation from boride inventory and seed consumption. The $N_2$ carrier is another potential nitrogen source; unlabeled N cannot be assigned uniquely to the seed without an independently labeled carrier or an Ar-carrier comparison. For NanoSIMS, topography, charging, sputter mixing and counting statistics must be measured for the wire geometry. [S44] The diamond atom-probe study used conductive boron-doped material and laser-assisted evaporation; reconstructed depth was typically 10–30 nm before specimen failure, with substantial depth-scale uncertainty. [S45] It supplies a method precedent, not guaranteed performance on an insulating cBN wire.

Senga and colleagues used dark-field vibrational STEM–EELS with a probe approximately 0.2 nm wide and measured an approximately 8 meV optical-phonon shift between $^{12}C$ and $^{13}C$ graphene. [S46] *A* spectrum averaged contributions from roughly four carbon atoms; the experiment did not identify each atom's isotope independently. In hBN, Li and colleagues reported mode-dependent phonon transition widths of approximately 0.5–3.3 nm across an abrupt isotope interface, despite a probe narrower than 0.1 nm. [S47] Thus a phonon-energy profile need not reproduce an isotope-composition boundary. Reference specimens of the same phase, thickness and orientation would be needed to assess spatial resolution and separate isotope shifts from strain or structural changes in a diamond or cBN wire.

The isotope signal and structural measurement should be registered to the same wire segment. Diffraction and bonding contrast can distinguish diamond/cBN from graphite/hBN or amorphous material, but ordinary TEM imaging does not itself resolve isotope identity. Where successive segments are shorter than the usable isotope-mapping resolution, first use longer labeling intervals, a longer extension or a larger-diameter specimen.

## S5 Measurements of the resulting filaments

Tensile or bending measurements should report crystal orientation, gauge length, cross-section and the location of fracture. Diamond {111} cleavage and surface flaws may dominate a single crystal, while grain boundaries, pores and interfaces can govern a polycrystalline filament. The reported diamond-needle bending and cBN-pillar compression results motivate these measurements but concern different load geometries. [S48, S49, S50]

Where diffusive transport applies, measurements at several lengths with comparable cross-section and microstructure can separate wire resistance from end-contact resistance. Dense wires must be distinguished from particle-assembled arrays, for which interfaces and packing can dominate. [S51] Diameter, orientation, grain structure and isotope composition are independently relevant to thermal transport. [S52, S53, S2] For scanning sensing, optical brightness, magnetic-resonance contrast and coherence should be measured on the grown wire rather than inferred from bulk crystals or powders.